\documentclass[preprintnumbers,prd,twocolumn,showpacs,
nofootinbib,
superscriptaddress]{revtex4}
 \usepackage[utf8x]{inputenc}
\usepackage{graphicx}
\usepackage{epsfig}
\usepackage{bm}
\usepackage{amssymb}
\usepackage{float}
\usepackage{amsmath}
\usepackage{dcolumn}
\usepackage{cancel}
\usepackage[colorlinks]{hyperref}
\usepackage[usenames,dvipsnames]{color}
\hypersetup{
     breaklinks=true,
    pdfstartview={FitH},    % fits the width of the page to the window
    colorlinks=true,       % false: boxed links; true: colored links
    linkcolor=blue,          % color of internal links
    citecolor=red,        % color of links to bibliography
    filecolor=magenta,      % color of file links
    urlcolor=blue,           % color of external links
    anchorcolor=green,      % Color for anchor text
    linktocpage=true
}

\begin{document}

\title{Cosmological applications of Myrzakulov gravity }

\author{Emmanuel N. Saridakis}
\email{msaridak@phys.uoa.gr}
\affiliation{Chongqing University of Posts \& Telecommunications, Chongqing, 
400065, 
China}
\affiliation{Department of Physics, National Technical University of Athens, 
Zografou
Campus GR 157 73, Athens, Greece}
\affiliation{National Observatory of Athens, Lofos Nymfon, 11852 Athens, 
Greece}
\affiliation{ Eurasian  International Center for Theoretical
Physics, Eurasian National University, Nur-Sultan 010008, Kazakhstan}

 \author{Shynaray Myrzakul}
\email{srmyrzakul@gmail.com}
\affiliation{ Eurasian  International Center for Theoretical
Physics, Eurasian National University, Nur-Sultan 010008, Kazakhstan}

 \author{Kairat Myrzakulov}
\affiliation{ Eurasian  International Center for Theoretical
Physics, Eurasian National University, Nur-Sultan 010008, Kazakhstan}

\author{Koblandy Yerzhanov} \email{yerzhanovkk@gmail.com} 
\affiliation{ Eurasian  International Center for Theoretical
Physics, Eurasian National University, Nur-Sultan 010008, Kazakhstan}

\begin{abstract}  
 We investigate the cosmological applications of Myrzakulov $F(R,T)$ gravity. 
In this theory ones uses a specific but non-special connection, and thus both 
curvature and torsion are dynamical fields related to gravity. We introduce a 
parametrization that quantifies the deviation of curvature and torsion scalars 
form their corresponding values  obtained using the special Levi-Civita 
and Weitzenb{\"{o}}ck  connections, and we extract the cosmological field 
equations   following the mini-super-space procedure. Even for the simple case 
 where the action of the theory is linear in $R$ and $T$, we find that the 
Friedmann equations contain new terms of geometrical origin,  reflecting the 
non-special connection.  Applying the theory at late times we find that we can 
acquire the thermal history of the universe, where  dark energy can be  
quintessence-like or phantom-like, or behave exactly as a cosmological constant 
and thus reproducing   $\Lambda$CDM cosmology. Furthermore, we show that 
these features are obtained for other Lagrangian choices, too. Finally,
early-time application  leads to the de Sitter solution, as well as to an 
inflationary realization with the desired scale-factor evolution.

\end{abstract}

\pacs{98.80.-k,  95.36.+x, 04.50.Kd}

%---------------------------------------------------
\maketitle
%---------------------------------------------------

\section{Introduction}

According to the standard paradigm of cosmology, which is based on an 
increasing amount of observational data, the universe went through two phases 
of accelerating expansion at early and late times of the cosmological evolution.
Although for   late time acceleration the cosmological constant may be the 
best explanation, the possibility that the acceleration has a dynamical nature, 
as well as some possible tensions, may ask for a modification of our 
knowledge, something which is definitely needed for the early time 
acceleration. There are two main ways that one could follow to achieve this. 
The first is to construct extended gravitational theories, which possess 
general relativity as a particular limit, but that in general can provide 
extra degrees of freedom that can describe the universe evolution successfully  
\cite{Capozziello:2011et,Nojiri:2010wj}. The second way is to consider the 
validity of general relativity and alter the standard model of particle 
physics, namely assume that there is extra matter content in the universe, 
such as the  dark energy \cite{Copeland:2006wr,Cai:2009zp} and/or the inflaton 
fields \cite{Bartolo:2004if}. Note that the first way has the additional 
theoretical advantage that it may lead to an improved renormalizable behavior 
\cite{Stelle:1976gc,Biswas:2011ar}.

In order to construct gravitational modifications one may start from the 
Einstein-Hilbert action, namely from the curvature description of gravity, and 
extend it suitably, such as in $F(R)$ gravity 
\cite{DeFelice:2010aj}, in $F(G)$ gravity   
\cite{Nojiri:2005jg,DeFelice:2008wz}, in Lovelock gravity
\cite{Lovelock:1971yv,Deruelle:1989fj}, etc. Alternatively, he can start from 
the equivalent, teleparallel formulation of gravity in terms of torsion 
\cite{Pereira,Maluf:2013gaa} and consider torsional modified gravities, such as 
$F(T)$ gravity \cite{Cai:2015emx,Ferraro:2006jd,Linder:2010py}, $F(T,T_G)$ 
gravity \cite{Kofinas:2014owa}, etc. Moreover, one could use non-metricity  as 
a way to construct gravitational modifications \cite{Harko:2018gxr}. 
Additionally, an  interesting class of modified gravity may arise by modifying 
the underlying geometry itself, considering for instance Finsler or 
Finsler-like geometries \cite{Bogoslovsky:1999pp,Mavromatos:2010jt,kour-stath-st 
2012,Basilakos:2013hua}. One of the interesting features of Finsler framework 
is that the role of the non-linear connection may bring the extra degrees of 
freedom that then make this gravitational modification phenomenologically 
interesting \cite{Triantafyllopoulos:2018bli,Ikeda:2019ckp}, and this feature 
was also obtained through the different theoretical framework of metric-affine 
theories \cite{Hehl:1994ue,BeltranJimenez:2012sz,Tamanini:2012mi}.

One interesting gravitational modification was obtained by R. Myrzakulov in 
\cite{Myrzakulov:2012qp}, namely the $F(R,T)$ gravity. In this theory ones uses 
a specific but non-special connection, and thus both curvature and torsion are 
dynamical fields related to gravity.\footnote{In 
\cite{Myrzakulov:2012qp}, as well as in subsequent relevant 
works in the literature, the theory proposed by Myrzakulov was called $F(R,T)$ 
gravity. However, due 
to the fact that the same name was later used in \cite{Harko:2011kv} and 
subsequent works to denote a completely different theory (in which $T$ is the 
trace of the energy-momentum tensor), in this work we prefer to call the theory 
at hand as  Myrzakulov gravity.} Hence, the theory has extra degrees of freedom 
coming from the non-special connection, as well as   extra degrees of freedom 
arising from the arbitrary function in the Lagrangian. The 
theory lies within the class of Riemann-Cartan theories, which in turn lie 
within the general family of affinely connected metric theories 
\cite{Conroy:2017yln}.
The investigation of some applications of the theory was performed 
in \cite{Myrzakulov:2012qp,Sharif:2012gz,Momeni:2011am,Capozziello:2014bna,
Feola:2019zqg}.  In particular, in \cite{Myrzakulov:2012qp}  some 
theoretical issues were discussed,  in  \cite{Sharif:2012gz} the energy 
conditions were investigated, in  \cite{Momeni:2011am}  the analysis 
focused on 
the theoretical relation with different scenarios, in 
\cite{Capozziello:2014bna} 
the Noether symmetries were explored, while in  \cite{Feola:2019zqg} the 
authors focused  on  neutron stars.

In the present 
work we are interested in studying the resulting cosmology of such a 
framework in a realistic way, calculating additionally the evolution of 
observables quantities  such as the density parameters and the effective dark 
energy equation-of-state parameter.  We will  express the theory as a 
deformation from both general 
relativity and its teleparallel equivalent and thus by applying the 
mini-super-space 
approach we can investigate the cosmological behavior focusing on the effect of 
the connection.

The plan of the work in the following. In Section \ref{Themodel} we review  
Myrzakulov gravity and we extract the cosmological equations. In Section 
\ref{applications} we study the cosmological applications  of various particular 
sub-cases, at late and early 
times. Finally, in Section 
 \ref{Conclusions} we summarize our results.

%%%%%%%%%%%%%%%%%%%%%%%%%%%%%%%%%%%%%%%%%%%%%%%%%%%%%%%%%%%
\section{Myrzakulov gravity}
\label{Themodel}
%%%%%%%%%%%%%%%%%%%%%%%%%%%%%%%%%%%%%%%%%%%%%%%%%%%%%%%%%%

In this section we briefly review Myrzakulov gravity based on 
\cite{Myrzakulov:2012qp}. As dynamical variables we use the  
tetrad field $e_a(x^\mu)$, as well as the 
  connection 1-forms $\omega^a_{\,\,\,
b}(x^\mu)$ which is used to define the parallel transportation. These fields 
can be expressed in components in terms of  
coordinates  as
$e_a=e^{\,\,\, \mu}_a\partial_\mu$ and $\omega^a_{\,\,\,b}=\omega^a_{\,\,\,
b\mu}dx^\mu=\omega^a_{\,\,\,bc}e^c$, while the dual tetrad is defined as 
$e^a=e^a_{\,\,\, \mu}d x^\mu$ \cite{Kofinas:2014owa}. Moreover, using the metric 
tensor   allows 
us to make the vielbein orthonormal and extract the relation 
\begin{equation}
\label{metrdef}
g_{\mu\nu} =\eta_{ab}\, e^a_{\,\,\,\mu}  \, e^b_{\,\,\,\nu},
\end{equation}
where
$\eta_{ab}=\text{diag}(-1,1,...1)$ is the Minkowski metric which is also 
used for raising/lowering the indices $a,b,...\,$. In the following  we
impose zero non-metricity, namely $\eta_{ab|c}=0$ ($|$ denotes the covariant 
differentiation with respect to $\omega^{a}_{\,\,\,bc}$), which implies that
$\omega_{abc}=-\omega_{bac}$.

Using the general connection $\omega^{a}_{\,\,\,bc}$ we can define the 
 curvature tensor, expressed in mixed (i.e. coordinate and tangent) components
 as \cite{Kofinas:2014owa}
\begin{eqnarray}
&&\!\!\!\!\!\!\!\!\! R^{a}_{\,\,\, b\mu\nu}\!=\omega^{a}_{\,\,\,b\nu,\mu}-
\omega^{a}_{\,\,\,b\mu,\nu}
+\omega^{a}_{\,\,\,c\mu}\omega^{c}_{\,\,\,b\nu}-\omega^{a}_{\,\,\,c\nu}
\omega^{c}_{\,\,\,b\mu}\,,
\label{curvaturebastard}
\end{eqnarray}
and the torsion tensor as 
\begin{equation}
T^{a}_{\,\,\,\mu\nu}=
e^{a}_{\,\,\,\nu,\mu}-e^{a}_{\,\,\,\mu,\nu}+\omega^{a}_{\,\,\,b\mu}e^{b}_{\,\,\,
\nu}
-\omega^{a}_{\,\,\,b\nu}e^{b}_{\,\,\,\mu}\,,
\label{torsionbastard}
\end{equation}
where comma denotes differentiation.
As one can observe, the torsion tensor depends on both the
tetrad and the connection, however  the
curvature tensor depends only on the connection.

Amongst the infinite choices of connections, the Levi-Civita   
$\Gamma_{abc}$ is the only one that leads to 
 vanishing torsion, while for clarity we denote its curvature (Riemann) tensor 
with the index  ``LC'', namely $R^{(LC)a}_{\,\,\,\ \ \ \ \ \, 
b\mu\nu}=\Gamma^{a}_{\,\,\,b\nu,\mu}-
\Gamma^{a}_{\,\,\,b\mu,\nu}
+\Gamma^{a}_{\,\,\,c\mu}\Gamma^{c}_{\,\,\,b\nu}-\Gamma^{a}_{\,\,\,c\nu}
\Gamma^{c}_{\,\,\,b\mu}$.
Hence, a general connection  is   related to the
Levi-Civita one  through  
\begin{equation}
\omega_{abc}=\Gamma_{abc}+\mathcal{K}_{abc}\,,
\label{omega}
\end{equation}
where $
\mathcal{K}_{abc}=\frac{1}{2}(T_{cab}-T_{bca}-T_{abc}
)=-\mathcal{K}_{bac}$ is the contorsion tensor. 
On the other hand the Weitzenb{\"{o}}ck connection  
$W_{\,\,\,\mu\nu}^{\lambda}$ leads to vanishing curvature, and in
all coordinate frames is
 defined in terms of the tetrad as
\begin{eqnarray}
W_{\,\,\,\mu\nu}^{\lambda}=e_{a}^{\,\,\,\lambda}e^{a}_{\,\,\,\mu
, \nu } .
\label{Weinzdef}
\end{eqnarray}
Note that due to its inhomogeneous transformation law the Weitzenb{\"{o}}ck 
connection expressed in tangent-space
components is simply $W_{\,\,\,bc}^{a}=0$, while it leads to 
$e_{a\,\,\,\,|\nu}^{\,\,\,\mu}=0$, i.e. the tetrad is autoparallel with respect 
to the Weitzenb{\"{o}}ck 
connection \cite{Kofinas:2014owa}. Hence, the corresponding torsion tensor 
(denoting by the index ``W'') is 
$T^{(W)\lambda}_{\,\,\,\ \ \ \ \ \mu\nu}=W^{\lambda}_{\,\,\,\nu\mu}-
W^{\lambda}_{\,\,\,\mu\nu}\,.$
 
In general relativity (GR) ones imposes the Levi-Civita connection, and thus 
all information of the gravitational field is embedded in the corresponding 
curvature (Riemann) tensor, whose contractions leads to the Lagrangian of the 
theory, namely the Ricci scalar $R^{(LC)}$ given by 
\begin{eqnarray}
 &&
 \!\!\!\!\!\!\!\!\!\!\!\!\!\!\!\!\!\!\!\!\!\!\!\!\!\!\!\!\!\!
 R^{(LC)}=\eta^{ab} e^{\,\,\, \mu}_a e^{\,\,\, \nu}_b  \left[
 \Gamma^{\lambda}_{\,\,\,\mu\nu,\lambda}
-
 \Gamma^{\lambda}_{\,\,\,\mu\lambda,\nu}\right.\nonumber\\
&&\ \ \ \ \  \ \ \ \ \ \left.
 + \Gamma^{\rho}_{\,\,\,\mu\nu}\Gamma^{\lambda}_{\,\,\,\lambda\rho}
-\Gamma^{\rho}_{\,\,\,\mu\lambda}\Gamma^{\lambda}_{\,\,\,\nu\rho}
  \right].
\end{eqnarray}
%{\small{
%\begin{equation}
% R^{(LC)}=\eta^{ab} e^{\,\,\, \mu}_a e^{\,\,\, \nu}_b  \left[
 %\Gamma^{\lambda}_{\,\,\,\mu\nu,\lambda}
%\!-\!
% \Gamma^{\lambda}_{\,\,\,\mu\lambda,\nu}
% \!+\! \Gamma^{\rho}_{\,\,\,\mu\nu}\Gamma^{\lambda}_{\,\,\,\lambda\rho}
% \!-\!\Gamma^{\rho}_{\,\,\,\mu\lambda}\Gamma^{\lambda}_{\,\,\,\nu\rho}
%  \right].
%\end{equation}}}
Moreover, in curvature modified 
gravity one uses $R^{(LC)}$ in order to construct various extensions, such as 
in $F(R)$ gravity \cite{DeFelice:2010aj}.
On the other hand, in teleparallel 
equivalent of general relativity (TEGR) one imposes the  Weitzenb{\"{o}}ck 
connection, and therefore the gravitational field is described by the 
corresponding torsion tensor, whose contractions leads to the Lagrangian of 
the theory, namely the torsion scalar $T^{(W)}$ given as \cite{Kofinas:2014owa}
\begin{eqnarray} 
&&
\!\!\!\!\!\!\!\!\!\!
T^{(W)}=\frac{1}{4}
\left(W^{\mu\lambda\nu}-
W^{\mu\nu\lambda} \right)
\left(W_{\mu\lambda\nu}  -W_{\mu\nu\lambda}\right)\nonumber\\
&&\ \ \ \ \ 
+\frac{1}{2} \left(W^{
\mu\lambda\nu }
-W^{
\mu\nu\lambda } \right)
\left(W_{\lambda\mu\nu}
-W_{\lambda\nu\mu}\right)
\nonumber\\
&&\ \ \ \ \ 
- \left(  
W_{\nu}^{\,\,\,\mu\nu}
-W_{\nu}^{\,\,\,\nu\mu}\right)  
\left( W^{\lambda}_{\,\,\,\mu\lambda}-W^{\lambda}_{\,\,\,\lambda\mu}\right).
\label{TdefW}
 \end{eqnarray}
Furthermore, in torsional modified 
gravity one uses $T^{(W)}$ in order to construct modifications, such as 
in $F(T)$ gravity  \cite{Cai:2015emx}.

As it is well known, both GR and TEGR possess two dynamical degrees of freedom, 
and thus describing a massless spin-two field, which lies in the core of 
standard gravitational description. On the other hand, in general a modified 
gravity possesses more degrees of freedom, whose effects on cosmological 
evolution, as well as on the theoretical properties of the theory, are exactly 
the motivation of constructing such 
modifications. As we mentioned in the 
introduction, one way that is known to introduce extra degrees of freedom is 
the 
consideration of non-special connections, namely going beyond the  
Levi-Civita and  
Weitzenb{\"{o}}ck ones. Thus, if ones uses a connection that 
has both non-zero curvature and torsion, the resulting theory will in general 
possess extra degrees of freedom, even if the imposed Lagrangian is simple, 
while if the imposed Lagrangian is complicated this will introduce even more 
degrees of freedom. In \cite{Myrzakulov:2012qp} Myrzakulov constructed $F(R,T)$ 
modified gravity, namely he let the connection to be non-special and thus 
having both non-zero curvature and non-zero torsion, while in the Lagrangian a 
general function of the curvature and torsion scalars was also allowed. Thus, 
the action of such a theory is  
\begin{equation}
S = \int d^{4}x e \left[ \frac{F(R,T)}{2\kappa^{2}}   +L_m \right],
\label{action1}
\end{equation}
with $e = \text{det}(e_{\mu}^a) = \sqrt{-g}$,  $\kappa^2=8\pi G$   the 
gravitational constant, and where for completeness we have also introduced 
the matter Lagrangian $L_m$.  We mention that in the above arbitrary function 
$F(R,T)$, the $R$ and $T$ are the curvature and torsion scalars corresponding 
to the imposed non-special connection, which are easily calculated as 
\cite{Kofinas:2014owa}
\begin{eqnarray}
&&
 T=\frac{1}{4}T^{\mu\nu\lambda}T_{\mu\nu\lambda}+\frac{1}{2}T^{\mu\nu\lambda}
T_{\lambda\nu\mu}-T_{\nu}^{\,\,\,\nu\mu}T^{\lambda}_{\,\,\,\lambda\mu},
\label{Tdef2}
\\
&&
 R=R^{(LC)}+T-2T_{\nu\,\,\,\,\,\,\,\,;\mu}^{\,\,\,\nu\mu}\,,
 \label{Rdef2}
 \end{eqnarray}
with $;$ denoting covariant differentiation with respect to the 
Levi-Civita connection. As one can see, $T$ depends on the tetrad, its first 
derivative and the connection, while $R$ depends on the tetrad, its first 
derivative, the connection and its first derivative, however its acquires a 
dependence on the second derivative of the tetrad due to the last term of 
(\ref{Rdef2}). Hence, as it has been extensively  discussed in the torsional 
literature \cite{Cai:2015emx}, if $R$ is used linearly in the Lagrangian the 
last term of (\ref{Rdef2}) will be a total derivative and therefore the 
resulting field equations will not contain higher derivatives, however in the 
general case they will do.
Finally, observing the forms of (\ref{TdefW}),(\ref{Tdef2}),(\ref{Rdef2}) we 
deduce that
we can  write  
\begin{eqnarray}
&&T=T^{(W)}+v, 
\label{T1}
\\
&&
R=R^{(LC)} + u, 
\label{R1}
\end{eqnarray}
where $v$ is a scalar quantity depending   on the tetrad, its first 
derivative and the connection, and $u$ is a scalar quantity depending on 
the   tetrad, its first and second derivatives, and the connection 
and its first derivative. Hence, $u$ and $v$ quantify the information on the 
specific imposed connection. 

 We mention here that  in our formulation we prefer to follow a 
short of effective approach, where we parametrize the connection structure in 
the way it appears at the level of scalars and not at the initial level of 
tensors, keeping of course in mind the general features (e.g. knowing the 
maximum order of derivatives appearing in the general tensors allows us to know 
the maximum number of derivatives in the scalars). 
Definitely, we could have 
formulated the theory by imposing the parametrization of the connection 
straightaway at the general symbolic level of tensors and connections, however 
proceeding to the calculation of scalars, of the action and then of the field 
equation, would result to increased complexity that would make the model 
difficult and inconvenient to be used for cosmological purposes.  
 
In the case where the imposed connection is the Levi-Civita 
one, then $u=0$ and $v=-T^{(W)}$, and thus the above theory coincides with the 
usual $F(R)$ gravity, which  then coincides with GR by choosing $F(R)=R$. 
Additionally, in the case where the imposed connection is the Weitzenb{\"{o}}ck 
one, then $v=0$ and $u=-R^{(LC)}$ and therefore it coincides with $F(T)$ 
gravity,  and it then coincides with TEGR by choosing $F(T)=T$.

In order to extract the field equations we have to choose a specific 
connection and then perform variation of the action (\ref{action1}) in terms 
of the tetrad field. We stress here that the underlying connection is a 
non-special one, however still it is a specific one in a given application of 
the theory, and thus one does not need to consider a variation in terms of the 
connection. Nevertheless, in order to avoid complications, and moreover to 
exploit in the most efficient  way the parametrization through the deformation 
functions 
$u$ and $v$, we will   apply a mini-super-space variation.

We consider explicitly the homogeneous and isotropic  flat 
Friedmann-Robertson-Walker (FRW) geometry 
$
ds^2= dt^2-a^2(t)\,  \delta_{ij} dx^i dx^j,
$
which corresponds to the tetrad choice 
$e^a_{\,\,\,\mu}={\rm
diag}[1,a(t),a(t),a(t)]$,
with $a(t)$ the scale factor.
In this case it is known that
\begin{eqnarray}
&&R^{(LC)}=6   \left( \frac{\ddot{a}}{a}+  \frac{\dot{a}^{2}}{a^2}\right), 
\nonumber\\
&&T^{(W)}=-6 \left(  \frac{\dot{a}^{2}}{a^2} \right).
\label{4.2}
\end{eqnarray}
Additionally, as usual in mini-super-space approach, we need to replace $L_m$ 
by the specific function of the matter energy density 
\cite{Paliathanasis:2014iva,Paliathanasis:2015aos,Dimakis:2016mip}, and since 
we are focusing on the 
   FRW geometry  we have $L_m=-\rho_m(a)$. Finally, as we 
mentioned after 
(\ref{T1}),(\ref{R1}), $v$ is a function  
of the tetrad, its first 
derivative and the connection, while $u$ is a function
of the tetrad, its first and second derivatives, and the connection 
and its first derivative. Hence, in the FRW geometry we can consider the 
general case that $u=u(a,\dot{a},\ddot{a})$ and $v=v(a,\dot{a})$.

In order to proceed we have to examine  the linear $F(R,T)$ case separately 
from the general one, since in the later an additional step of conformal 
transformation is needed.

\subsection{$F(R,T)=R+\lambda T$}

As a first sub-case we consider the simple case where the action is linear in 
both $R$ and $T$, namely $F(R,T)=R+\lambda T$ (the coupling coefficient of $R$ 
can be absorbed into $\kappa^2$ and thus we omit it).  Note that these 
scalars are given by (\ref{T1}),(\ref{R1}) and correspond to the specific but 
non-special connection, hence they are neither the simple Levi-Civita nor the  
simple Weitzenb{\"{o}}ck ones.  
Inserting the above mini-super-space expressions, as well as 
(\ref{T1}),(\ref{R1}), 
into the action (\ref{action1}), we obtain $S=\int Ldt$, with 
\begin{eqnarray}
&&
\!\!\!\!\!\!\!\!\!\!\!
L= 
\frac{3}{\kappa^2}\left[\lambda+1\right]a\dot{a}^{2}-
\frac{ a^{3}}{2\kappa^2}\left[ u(a,\dot{a},\ddot{a})+\lambda
v(a,\dot{a}) \right]\nonumber\\
&& \,
+ a^3  \rho_m(a)
.\label{4.2}
\end{eqnarray}
Extracting the equations of motion for $a$, alongside   
the Hamiltonian constraint  $
{\cal H}=\dot{a}\left[\frac{\partial L}{\partial 
\dot{a}}- \frac{\partial}{\partial t}\frac{\partial L}{\partial 
\ddot{a}}\right]+\ddot{a}\left(\frac{\partial L}{\partial 
\ddot{a}}\right)-L=0 $, we finally acquire:
\begin{eqnarray}
&&
\!\!\!\!\!\!\!\!\!\!\!\!\!\!\!
3H^{2}=
\frac{\kappa^2  }{\left(1+\lambda\right) } \rho_m\nonumber\\
&&\ \ 
+\frac{1}{\left(1+\lambda\right)}\left[
\frac{Ha}{2} \left(u_{\dot{a}}+v_{\dot{a}} \lambda-\dot{u}_{\ddot{a}}\right) 
-\frac{1}{2} 
(u+\lambda v)
\right.\nonumber\\
&&
 \ \ \ \ \ \ \ \ \ \ \ \ \ \ \ \ 
\left.
+
\frac{a u_{\ddot{a}}}{2}  \left(\dot{H}-2 H^2\right)    
\right],
\label{eq1b}
\end{eqnarray}
\begin{eqnarray}
&&
\!\!\!\!\!\!\!\!\!\!\!\!\!\!\!
2\dot{H}+3H^2
=-\frac{\kappa^2 }{\left(1+\lambda\right)}p_m
\nonumber\\
&&\ \ \ \ \ \ \ \ \ \  
+\frac{1}{\left(1+\lambda\right)}\!
\Big[ \frac{Ha}{2} \left(u_{\dot{a}}+v_{\dot{a}} 
\lambda\right)
 -\frac{1}{2} (u+ \lambda v) \nonumber\\
&&\ \ \ \ \ \ \ \ \ \  
 +\frac{a}{6} 
\left(-u_a-\lambda v_a+{\dot{u}_{\dot{a}}}+\lambda 
{\dot{v}_{\dot{a}}}\right)\nonumber\\
&&\ \ \ \ \ \ \ \ \ \ 
-\frac{a}{2}\left(\dot{H}+3H^2\right)u_{\ddot{a}}-H a 
\dot{u}_{\ddot{a}}-\frac{a}{6} 
\ddot{u}_{\ddot{a}}
\Big] ,
\label{eq2b}
\end{eqnarray}
with $H=\frac{\dot{a}}{a}$ is the Hubble parameter and where the subscripts 
$a,\dot{a}$, denote partial derivatives with respect to this 
argument. Additionally, $p_m$ denotes the matter pressure whose conservation 
equation is 
\begin{eqnarray}
 \dot{\rho}_m+3H(\rho_m+p_m)=0.
 \label{matterconserv}
\end{eqnarray}
As we observe, in the Friedmann equations we obtain new terms of geometrical 
origin, namely terms that reflect the new degrees of freedom brought by  the 
non-special connection through the parametrization in terms of $u$ and $v$. 
Hence, due to the non-special connection, the theory is not trivial even if the 
Lagrangian has a simple form.
These terms will have interesting cosmological implications as we will see later 
on.

\subsection{General $F(R,T)$ case}

In the case where $F(R,T)$ is non-linear in $R$ and $T$ we need to perform an 
additional step of conformal transformations in order to be able to apply the 
mini-super-space procedure. 
Extending the analysis of \cite{Yang:2010ji,Wright:2016ayu,Gakis:2019rdd}  we 
start by 
introducing two scalar fields $\phi_{1}$ and $\phi_{2}$ and re-writing the 
action (\ref{action1}) as
\begin{eqnarray}
&&
\!\!\!\!\!\!\!\!\!\!\!\!\!\!\!\!\!\!\!
S= \frac{1}{2\kappa^2} \int d^4x e 
\left[F_{\phi_1}(\phi_1,\phi_2)(R-\phi_1)\right. 
\nonumber\\
&&
\ \ \ \ \ \ \ \ \ \ \ \ \
+F_{\phi_2}(\phi_1,\phi_2)(T-\phi_2) \nonumber\\
&&
\ \ \ \ \ \ \ \ \ \ \ \ \,
\left.
  +F(\phi_1,\phi_2)+2\kappa^{2}L_m\right], \label{action3.25}
\end{eqnarray}
with $F_{\phi_1}(\phi_1,\phi_2)=\partial F(\phi_1,\phi_2)/\partial\phi_1$, 
$F_{\phi_2}(\phi_1,\phi_2)=\partial F(\phi_1,\phi_2)/\partial\phi_2$ and where 
$F(\phi_1,\phi_2)$ has the same functional form with $F(R,T)$. We mention that 
in order for the above re-writing to be meaningful, $F(R,T)$ must be non-linear 
in $R$ and $T$. If it is linear in one of $R$ or $T$ then the corresponding 
scalar field disappears, and we transit to the case of the previous subsection. 
Introducing now 
the conformal transformation in terms of the tetrad field through 
\cite{Wright:2016ayu,Gakis:2019rdd}
\begin{eqnarray}
\hat{e}^a_{\,\,\,\mu}=\Omega(x)e^a_{\,\,\,\mu},
\end{eqnarray} 
where $\Omega(x)$ is continuous, non-vanishing, finite real function,
and choosing  $\Omega^{2}=F_{\phi_1}$, we can re-write action (\ref{action3.25})
 as  
 \begin{eqnarray}
&&
\!\!\!\!\!\!\!\!\!\!\!\!\!\!\!\!\!\!\!\!\!\!\! 
S=  \int d^4x e \Big[\frac{ R}{2\kappa^2}+\frac{f(\phi_{1},\phi_{2}) 
T}{2\kappa^2}   
\nonumber\\
&& \ \ \ \ \ \ 
+\frac{1}{2}  \partial^{\mu} 
\phi_{1} \partial_\mu \phi_{1}
+ \frac{1}{2}\partial^{\mu} 
\phi_{2} \partial_\mu \phi_{2} \nonumber\\
&& \ \ \ \ \ \
-V(\phi_{1},\phi_{2})+ 
F^{-2}_{\phi_1}L_m(F^{-1}e^a_{\,\,\,\mu})\Big],
\label{action2}
\end{eqnarray}
%\begin{eqnarray}
%&&
%\!\!\!\!\!\!\!\!\!\!\!\!\!\!\!\!
%S=  \int d^4x e \left[\frac{ R}{2\kappa^2}+\frac{f(\phi_{1},\phi_{2}) 
%T}{2\kappa^2} +\frac{1}{2}  \partial^{\mu} 
%\phi_{1} \partial_\mu \phi_{1}\right. 
%\nonumber\\
%&&\!\!\! 
%\left.
%+ \frac{1}{2}\partial^{\mu} 
%\phi_{2} \partial_\mu \phi_{2} 
%-V(\phi_{1},\phi_{2})+ 
%F^{-2}_{\phi_1}L_m(F^{-1}e^a_{\,\,\,\mu})\right],
%\label{action2}
%\end{eqnarray}
omitting the hats for simplicity, and 
where we have defined
\begin{eqnarray}
f(\phi_1,\phi_2)= 
\frac{F_{\phi_2}(\phi_1,\phi_2)}{F^{2}_{\phi_1}(\phi_1,\phi_2)},
\label{fdef}
\end{eqnarray}
and
\begin{equation}
V(\phi_1,\phi_2)=\phi_1 F_{\phi_1}(\phi_1,\phi_2)+\phi_2 
F_{\phi_2}(\phi_1,\phi_2)-F(\phi_1,\phi_2).\label{4.44} 
\end{equation} 

As in the previous subsection, we proceed by considering explicitly the 
homogeneous and isotropic  flat 
 FRW geometry. Inserting the mini-super-space expressions described in the 
previous subsection
into   action (\ref{action2}), we obtain $S=\int Ldt$, with 
\begin{eqnarray}
&&
\!\!\!\!\!\!\!\!
L= 
-\frac{3}{\kappa^2}\left[f(\phi_1,\phi_2)+1\right]a\dot{a}^{2}+   
a^3 \!\!\left[ \frac{\dot{\phi}^{2}_1} {2} 
+ \frac{ \dot{\phi}^{2}_{2}}{2}- V(\phi_1,\phi_2)\right]
\nonumber\\
&&+a^{3}\!\left[ \frac{u(a,\dot{a},\ddot{a})\!+\!f 
v(a,\dot{a})}{2\kappa^2}\right]-  a^3\rho_m(a) 
F^{-2}_{\phi_1}(\phi_1,\phi_2).\label{4.2}
\end{eqnarray}
Performing variation in terms of $a$,$\phi_1$,$\phi_2$, alongside the 
the Hamiltonian constraint
$
{\cal H}=\dot{a}\left[\frac{\partial L}{\partial 
\dot{a}}- \frac{\partial}{\partial t}\frac{\partial L}{\partial 
\ddot{a}}\right]+\ddot{a}\left(\frac{\partial L}{\partial 
\ddot{a}}\right)+\dot{\phi}_{1}\frac{\partial 
L}{\partial \dot{\phi}_{1}}+\dot{\phi}_{2}\frac{\partial L}{\partial 
\dot{\phi}_{2}}-L=0 $, we finally acquire:
\begin{eqnarray}
&&
3H^{2}=
\frac{\kappa^2}{\left(1+f\right)}\left[\frac{1}{2}\left( 
\dot{\phi}^{2}_{1}+ 
{\dot{\phi}}^{2}_{2}\right)+ V+ F^{-2}_{\phi_1} \rho_m
\right]
\nonumber\\
&&
\ \ \ \ \ \ \ \ \ \ 
+ 
\frac{1}{\left(1+f\right)}\left[
\frac{Ha}{2} \left(u_{\dot{a}}+v_{\dot{a}} f\right) -\frac{1}{2} 
(u+fv)
\right. \nonumber\\
&&
\ \ \ \ \ \ \ \ \ \ 
\ \ \ \ \ \ \ \ \ \ \ \ \ \
\left.
+\frac{au_{\ddot{a}}}{2} 
\left(\dot{H}-2 H^2\right)
-\frac{1}{2} 
\dot{u}_{\ddot{a}}\right],
\label{eq1}
\\
&&
\!\!\!\!\!\!\!\!\!\!\!
2\dot{H}+3H^2
= 
\frac{\kappa^2}{\left(1+f\right)}
\Big[\!-\!\frac{1}{2} 
\left(\dot{\phi}^{2}_{1}+\dot{\phi}^{2}_{2}\right)+ 
V\nonumber\\
&&
\ \ \ \ \ \ \ \ \ \ \ \ \ \ \ \ \ \ \ \ \ \ \ \
+ F^{-2}_{\phi_1} \Big(\rho_m + \frac{\dot{\rho}_m}{3H} 
  \Big) \Big]\nonumber\\
&&
 \ \ \ \ \ \ \   \ \   \ \  \
+ 
\frac{1}{\left(1\!+\!f\right)}
\Big[\frac{Ha}{2} \!\left(u_{\dot{a}}\!+\!v_{\dot{a}} 
f\right)
-2H\dot{f}-\frac{1}{2} (u\!+\!fv)\nonumber\\
&&
 \ \ \ \ \ \ \   \ \   \ \  \
+\frac{a}{6} 
\left(-u_a-fv_a+{\dot{u}_{\dot{a}}}+f {\dot{v}_{\dot{a}}}+\dot{f} 
v_{\dot{a}}\right)\nonumber\\
&&
 \ \ \ \ \ \ \   \ \   \ \  \
+\frac{au_{\ddot{a}}}{2} 
 \left(\dot{H}+3H^2\right)+a H 
\dot{u}_{\ddot{a}}+\frac{a}{6} \ddot{u}_{\ddot{a}}\Big] ,
\label{eq2}
\end{eqnarray}
\begin{eqnarray}
&&\ddot{\phi}_{1}=-3H\dot{\phi}_{1}-V_{\phi_{1}}-\frac{3}{\kappa^2 }H^2 
f_{\phi_{1}}\nonumber\\
&&  \ \ \ \ \ \ \ \, 
+\frac{v}{2 \kappa^2 } f_{\phi_{1}} +2 \rho_m
F^{-3}_{\phi_1}  F_{\phi_1\phi_1},
\label{eq3}
\\
&&\ddot{\phi}_{2}=
-3H\dot{\phi}_{2}-V_{\phi_{2}}-\frac{3}{\kappa^2 }H^2 
f_{\phi_{2}}\nonumber\\
&&  \ \ \ \ \ \ \ \, 
+\frac{v}{2 \kappa^2 } f_{\phi_{2}} +2    \rho_m
F^{-3}_{\phi_1} F_{\phi_1\phi_2},
\label{eq4}
\end{eqnarray}
 where the subscripts 
 $\phi_1$,$\phi_2$ denote partial derivatives with respect to this 
argument, and where we have used that $a(\rho_{m})_a=\dot{\rho}_m/H$. 
As one can straightforwardly check by differentiating (\ref{eq1}) and inserting 
  into (\ref{eq2}), eliminating the second time-derivatives of the scalar 
fields using (\ref{eq3}),(\ref{eq4}), the matter energy density is conserved, 
namely (\ref{matterconserv}) holds.

As we observe in the above equations we have two extra degrees of freedom, 
namely the two scalar fields, that arise from the conformal transformation of 
the $F(R,T)$ function considered in the action, as well as additional extra 
degrees of freedom arising form the non-special connection, and which have been 
quantified through $u$ and $v$. Hence, we deduce that Myrzakulov gravity is a 
very rich theory which may lead to big class of cosmological behaviors.
In the following section we will investigate particular sub-cases.

 We close this section by mentioning that in general each connection choice 
corresponds to a different class of theories, since  given 
a theory  for some choice of $F(R,T)$ and the connection,  it is not possible 
to reproduce the same theory for a different choice of $F(R,T)$ and a 
redefined  connection,  due to the fact that  
different connections give curvature and torsion scalars that have different 
   structure. As has been discussed in the literature (see 
e.g.\cite{Cai:2015emx}), the only possibility 
that one could acquire a  correspondence between various curvature and torsion 
theories, i.e between theories with different connections, is to use not only 
the simple curvature and torsion scalars, but all the higher-order   ones (such 
as the curvature Gauss-Bonnet, Lovelock, etc ones, and the corresponding 
torsional ones), nevertheless even this statement is still a 
conjecture  not a proven theorem.

\section{Cosmological applications}
\label{applications}

In this section we will study the cosmological applications of Myrzakulov 
gravity. As we mentioned above, in general the theory at hand possesses extra 
degrees of freedom arising from the arbitrary function $f(R,T)$ as well as from 
the non-special connection. Thus, in order to examine more economical 
scenarios, in the following we will examine some subcases separately.

\subsection{$F(R,T)=R+\lambda T$}

As a first sub-case we consider the simple model where the action is linear in 
both $R$ and $T$, namely $F(R,T)=R+\lambda T$. In this case the Friedmann 
equations are (\ref{eq1b}),(\ref{eq2b}).
These equations can be re-written in the standard form  
\begin{eqnarray}
3H^{2}&=& \kappa^2\left( 
\rho_m+\rho_{MG} \right)
\label{FR1a}
\\
2\dot{H}+3H^2
&=&  -\kappa^2 \left(p_m+ p_{MG}\right),
\label{FR2a}
\end{eqnarray}
where 
\begin{eqnarray}
&&
\!\!\!\!\!\!\!\!\!\!\!\!\!\!\!\!
\rho_{MG}=\frac{1}{\kappa^2} \Big[
\frac{Ha}{2} \left(u_{\dot{a}}+v_{\dot{a}} \lambda\right) -\frac{1}{2} 
(u+\lambda v)\nonumber\\
&&
\ \ \ \ \ \ \
+
\frac{a u_{\ddot{a}}}{2}  \left(\dot{H}-2 H^2\right)    
-3\lambda H^2\Big]
\label{rhoDEa1}\\
&&
\!\!\!\!\!\!\!\!\!\!\!\!\!\!\!\!
p_{MG}=
-\frac{1}{\kappa^2}
\Big[\frac{Ha}{2} \left(u_{\dot{a}}+v_{\dot{a}} 
\lambda\right)
 -\frac{1}{2} (u+ \lambda v)
 \nonumber\\
&&  \ \  \ \  \ \  \ \  \ 
 -\frac{a}{6} 
\left(u_a+\lambda v_a-{\dot{u}_{\dot{a}}}-\lambda 
{\dot{v}_{\dot{a}}}\right)\nonumber\\
&&  \ \  \ \  \ \  \ \  \ 
-\frac{a}{2}\left(\dot{H}+3H^2\right)u_{\ddot{a}}-H a 
\dot{u}_{\ddot{a}}
\nonumber\\
&&  \ \  \ \  \ \  \ \  \ 
-\frac{a}{6} 
\ddot{u}_{\ddot{a}}
-\lambda(2\dot{H}+3H^2)\Big].
\label{pDEa1}
\end{eqnarray}
As we observe, we 
obtain an effective sector arising from the non-special 
connection. 
As one can check, given the matter conservation equation (\ref{matterconserv}) 
we can obtain
\begin{eqnarray}
 \dot{\rho}_{MG}+3H(\rho_{MG}+p_{MG})=0,
\end{eqnarray}
and thus the effective sector is conserved. In summary, one can use this 
geometrical effective sector in order to study  the 
late-time universe, or to study inflation.

\subsubsection{Late-time acceleration}

Let us try to use the Friedmann equations (\ref{FR1a}),(\ref{FR2a}) in order to 
describe the late-time acceleration. First of all, as one can see, in the case 
where $\lambda=0$, i.e. where we consider a Lagrangian which is just the 
curvature $R$ corresponding to the non-special connection, and choosing a 
connection with $u=c_1 \dot{a}-c_2$, with $c_1$,$c_2$ constants, then we obtain 
\begin{equation}
 \rho_{MG}=-p_{MG}=\frac{c_2}{2\kappa^2}\equiv\Lambda. 
\end{equation}
Interestingly enough, 
we are able to re-obtain $\Lambda$CDM cosmology although we have not considered 
an explicit cosmological constant, namely the cosmological constant appears 
are a result of the structure of the underlying geometry. We mention that 
although for $\lambda=0$ we remove $T$ from the action, the geometry still has 
a non-special connection and thus a non-zero torsion. Such 
an effective appearance of the cosmological constant from the richer geometry 
is something that has been discussed in the literature for other geometrical 
modified gravities \cite{Ikeda:2019ckp,Minas:2019urp}, and reveals 
the capabilities of the theory.

We proceed by considering the $\lambda\neq0$ case. Choosing $u=c_1 \dot{a}-c_2$ 
and $v=c_3\dot{a}-c_4$   with $c_3$,$c_4$ constants   (these choices is 
just an example which has the advantage that the resulting 
cosmological equations are up to second order in time derivatives   and thus
free from Ostrogradsky ghosts, and hence   have a physical interest apart from 
a 
mathematical one), we obtain
\begin{eqnarray}
&&\rho_{MG}=\frac{1}{\kappa^2} \left[c-3\lambda H^2\right]
\label{rhoDEa}
\\
&&
p_{MG}=
-\frac{1}{\kappa^2}
\left[c-\lambda(2\dot{H}+3H^2)\right],
\label{pDEa}
\end{eqnarray}
with $c\equiv c_2+c_4$. Hence, in this scenario the geometrical sector 
constitutes an effective dark energy sector with the above energy density and 
pressure, and an equation-of-state parameter of the form
\begin{eqnarray}
&&w_{MG}\equiv w_{DE}=-1+\frac{2\lambda \dot{H}}{c-3\lambda H^2}.
\label{wDEa}
\end{eqnarray}
Interestingly enough, we can see that $w_{DE}$ can be both larger or smaller 
than -1, and thus the effective dark energy can be quintessence or phantom 
like.

\begin{figure}[ht]
\includegraphics[width=0.395\textwidth]{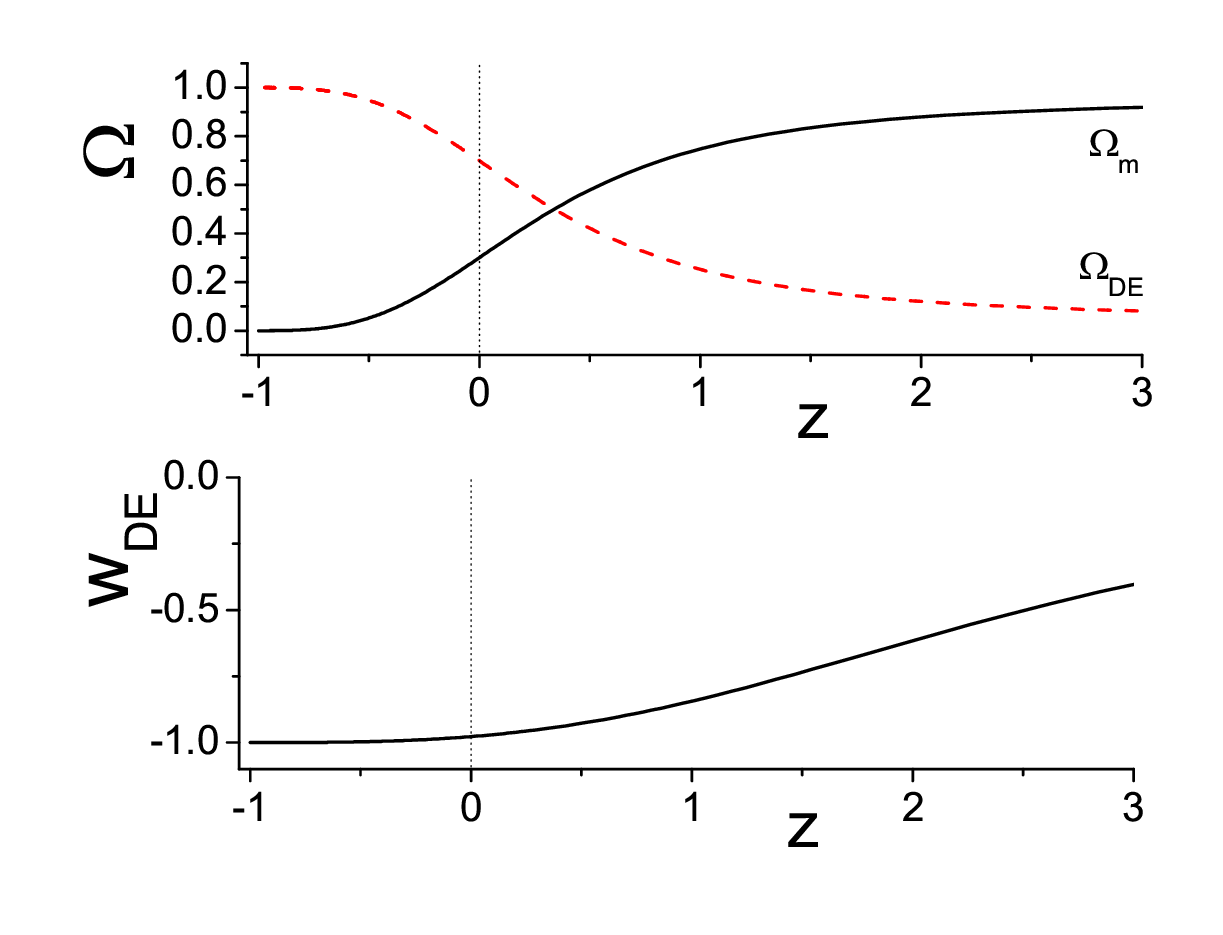}
\caption{
{\it{ Upper graph: The evolution of the matter density
parameter $\Omega_{m}$ (black-solid) and of 
effective dark energy
density parameter $\Omega_{DE}$ (red-dashed), as a function of the redshift 
$z$,
in the case of  $F(R,T)=R+\lambda T$ and $u=c_1 \dot{a}-c_2$ 
and $v=c_33\dot{a}-c_4$, for  $\lambda=-0.05$ and 
$c=c_2+c_4=1$,
in units where $\kappa^2=1$. We have imposed the initial conditions
$\Omega_{DE}(z=0)\equiv\Omega_{DE0}\approx0.69$  in agreement with observations.
Lower graph: The evolution of the corresponding effective dark
energy equation-of-state parameter $w_{DE}$ from (\ref{wDEa}). The dotted 
vertical line marks the present time.  
}} }
\label{Model1aomegas}
\end{figure}

In order to be more transparent we  solve the  Friedmann equations numerically 
for various values of the model parameters,
using as   independent variable  the
redshift    $
 z=a_0/a-1$, with the present scale factor set to $a_0=1$. For   
initial 
conditions we impose
$\Omega_{DE}(z=0)\equiv\Omega_{DE0}\approx0.69$  and thus
$\Omega_m(z=0)\equiv\Omega_{m0}\approx0.31$  as required by observations
\cite{Ade:2015xua}, where   $\Omega_i\equiv
\kappa\rho_i/(3H^2)$ are the density parameters of the corresponding fluid.   
In the
upper graph of Fig. \ref{Model1aomegas} we depict
$\Omega_{DE}(z)$ and $\Omega_{m}(z)$, while in the lower graph we present 
$w_{DE}(z)$. As we can see from 
 the upper graph, the scenario at hand can describe 
the standard
thermal history, namely the sequence of matter and dark energy eras, while in 
the future
($z\rightarrow-1$) the universe  results asymptotically  in a completely
dark-energy
dominated, de Sitter phase.  Moreover, from the lower graph of Fig. 
\ref{Model1aomegas} we observe
that the effective dark
energy equation-of-state parameter $w_{DE}$ lies in the quintessence 
regime and it quickly acquires the value $-1$.
 \begin{figure}[ht]
\includegraphics[width=0.395\textwidth]{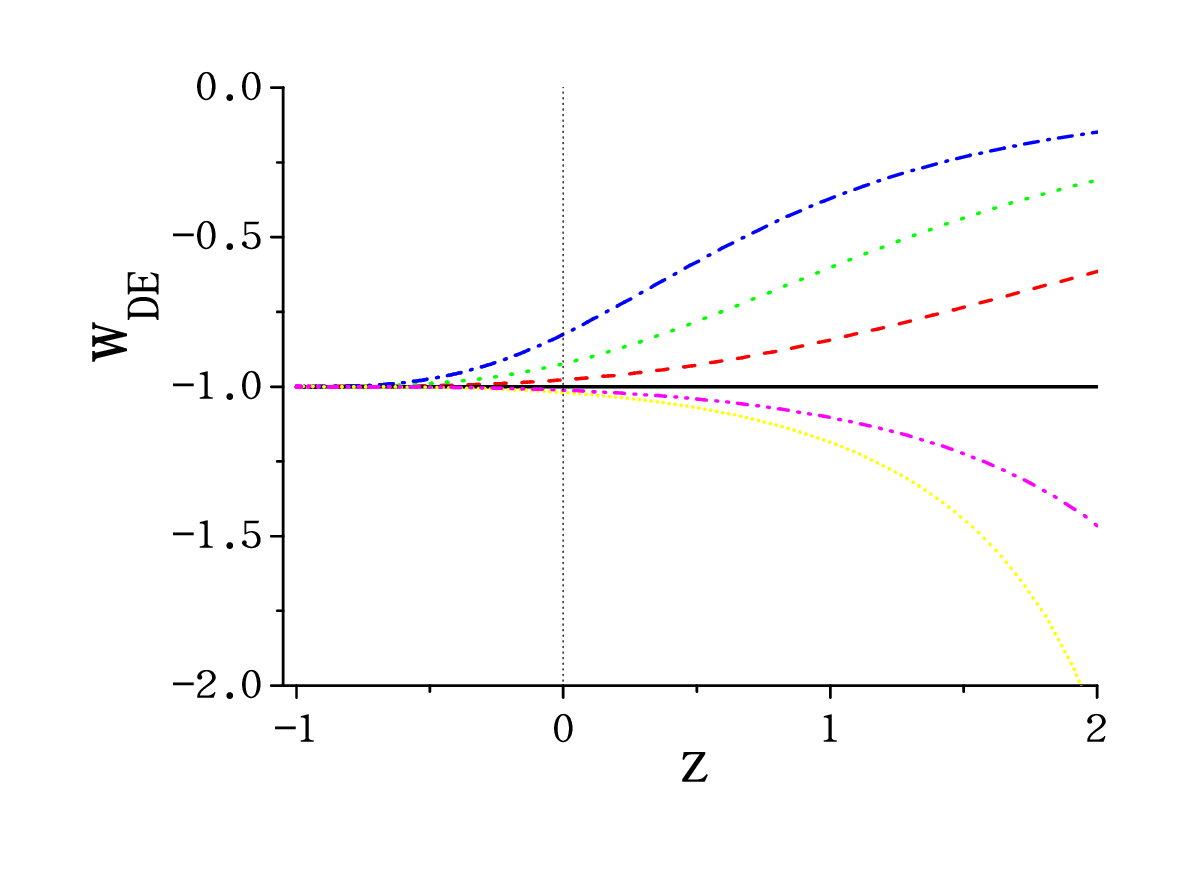}
\caption{
{\it{  The evolution of the equation-of-state parameter  $w_{DE}$  of the 
effective dark
energy    
in the case of  $F(R,T)=R+\lambda T$ and $u=c_1 \dot{a}-c_2$ 
and $v=c_3\dot{a}-c_4$,  for $c=c_2+c_4=1$ and various values of $\lambda$:
 $\lambda=-0.3$ (blue - dashed-dotted),  $\lambda=-0.2$ (green - dotted), 
  $\lambda=-0.05$ (red - dashed),  $\lambda=0$  (black - solid), $\lambda=0.03$ 
(magenta - dashed-dotted-dotted), $\lambda=0.05$ 
(yellow - short-dotted).
We have imposed the initial conditions
$\Omega_{DE}(z=0)\equiv\Omega_{DE0}\approx0.69$ and we use units where 
$\kappa^2=1$. The dotted vertical line marks the present time.
}} }
\label{Model1awde}
\end{figure}

In order to study the effect of the model parameters on  $w_{DE}$, in  
Fig. \ref{Model1awde} we
depict the evolution of  $w_{DE}(z)$ for various parameter choices. As we 
can see, the important parameter is $\lambda$, and in particular for 
$\lambda<0$ the effective dark energy is quintessence-like, while for 
$\lambda>0$ it is phantom-like. Finally, as we mentioned above, in the special 
case where $\lambda=0$ the effective dark energy behaves as a comsological 
constant, namely our scenario becomes exactly $\Lambda$CDM despite the fact 
that the underlying geometry and connection are not trivial.

As a second example in this sub-case let us consider  $u=c_1 
\frac{\dot{a}}{a}\ln\dot{a}$ 
and $v=s(a)\dot{a}$, with $s(a)$ an arbitrary function.  
These choices form a subclass  in which  the 
resulting 
cosmological equations are up to second order in time derivatives,  and 
moreover the dark energy energy density and pressure  do not depend on 
$a$,$\dot{a}$,$\ddot{a}$ but only on their combinations that give $H$ and 
$\dot{H}$. The former is a necessary requirement in order for the scenario to 
be free from 
Ostrogradsky ghosts, while the latter is imposed for simplicity. Clearly one 
can find large classes of choices that satisfy the above two requirements, and 
much wider classes if he abandons the second one. 
In this case 
(\ref{rhoDEa1}),(\ref{pDEa1}) give
\begin{eqnarray}
&&\rho_{MG}=\frac{1}{\kappa^2} \left[\frac{c_1}{2}H-3\lambda H^2\right]
\label{rhoDEb}
\\
&&
p_{MG}=
-\frac{1}{\kappa^2}
\left[\frac{c_1}{2}H+\frac{c_1}{6}\frac{\dot{H}}{H}-\lambda(2\dot{H}+3H^2)\right
] ,
\label{pDEb}
\end{eqnarray}
 while
\begin{eqnarray}
&&w_{MG}\equiv w_{DE}=-1+\frac{2\lambda 
\dot{H}-\frac{c_1}{6}\frac{\dot{H}}{H}}{\frac{c_1}{2}H-3\lambda H^2}.
\label{wDEb}
\end{eqnarray}
Similarly to the previous example, for this case too  $w_{DE}$ can be 
quintessence-like or 
phantom-like.

\begin{figure}[ht]
\includegraphics[width=0.395\textwidth]{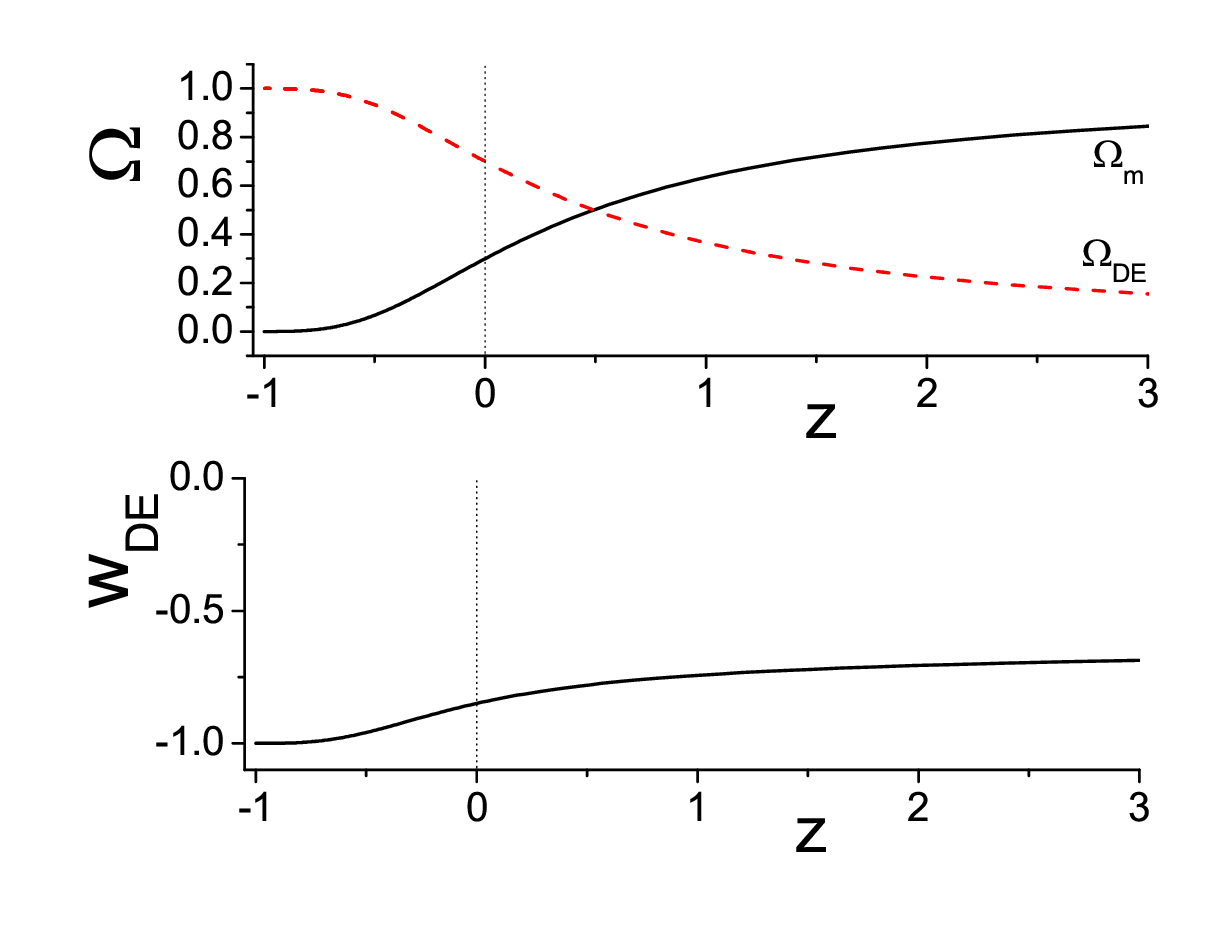}
\caption{
{\it{ Upper graph: The evolution of the matter density
parameter $\Omega_{m}$ (black-solid) and of 
effective dark energy
density parameter $\Omega_{DE}$ (red-dashed), as a function of the redshift 
$z$,
in the case of  $F(R,T)=R+\lambda T$ and  $u=c_1 
\frac{\dot{a}}{a}\ln\dot{a}$ 
and $v=s(a)\dot{a}$, for  $\lambda=-0.01$ and 
$c_1=4$,
in units where $\kappa^2=1$. We have imposed the initial conditions
$\Omega_{DE}(z=0)\equiv\Omega_{DE0}\approx0.69$  in agreement with observations.
Lower graph: The evolution of the corresponding effective dark
energy equation-of-state parameter $w_{DE}$ from (\ref{wDEb}). The dotted 
vertical line marks the present time.  
}} }
\label{Model2aomegas}
\end{figure}
We elaborate the cosmological equations numerically and in Fig. 
\ref{Model2aomegas} we present     the matter and dark energy density 
parameters, as well as the dark energy equation-of-state parameter.
 As we can see from  the upper graph, the scenario can describe 
the  sequence of matter and dark energy epochs, while in 
the future the universe  results asymptotically  in a completely
dark-energy dominated, de Sitter phase.  Additionally, from the lower graph of 
Fig. \ref{Model1aomegas} we observe that the effective dark
energy equation-of-state parameter $w_{DE}$ lies in the quintessence 
regime and it quickly acquires the value $-1$. However, the important feature 
of this scenario is that the above behavior is obtained although we have not 
considered neither an explicit cosmological constant, nor a constant term in 
the $u$ and $v$, namely it arises solely from the dynamical features of the 
non-special connection. This is an additional advantage of the scenario at hand.

\subsubsection{Inflation}

Let us try to use the Friedmann equations (\ref{FR1a}),(\ref{FR2a}) in order to 
obtain the realization of inflation. As usual, in the early universe we may 
neglect the matter sector. In this case, by choosing $u=c_1 \dot{a}-c_2$ 
and $v=c_3\dot{a}-c_4$ we obtain 
\begin{eqnarray}
3H^{2}&=&  c-3\lambda H^2 
\label{FR1infa}
\\
2\dot{H}+3H^2
&=&   c-\lambda(2\dot{H}+3H^2),
\label{FR2infa}
\end{eqnarray}
with $c\equiv c_2+c_4$. One can immediately see that the above equations accept 
the de Sitter solution $a(t)=e^{H_{dS}t}$, with 
$H_{dS}=\sqrt{\frac{c}{3(\lambda+1)}}$, which is the basis of any 
inflationary 
scenario. Once again we mention that the above  de Sitter solution is extracted 
without considering an explicit cosmological constant, namely it is of purely 
geometrical origin.

In order to acquire a  more realistic inflationary realization, 
with a 
successful exit after a desired e-folding, and particular desired values for 
the inflationary observables such as the scalar spectral index and the
tensor-to-scalar ratio, we can follow the reconstruction method, which is a 
well-applied method in standard potential-driven inflation. In particular, let 
us impose a specific $H(t)$, namely a specific scale factor $a(t)$, that 
provides the Hubble slow-roll parameters \cite{Martin:2013tda} and thus the 
inflationary observables we want.  We insert this $a(t)$ into the general 
Friedmann equations (\ref{FR1a}),(\ref{FR2a}) and as usual we neglect the 
 matter sector. We can now consider that the connection parametrization 
functions have simple forms with one argument that allow for the simple chain
rules to hold, namely if $u=u(a)$ then $u_{a}=\frac{\dot{u}}{\dot{a}}$ and 
$u_{\dot{a}}=0$,$u_{\ddot{a}}=0$, if $u=u(\dot{a})$ then 
$u_a=0$,$u_{\ddot{a}}=0$ and $u_{\dot{a}}=\frac{\dot{u}}{\ddot{a}}$, if 
$u=u(\ddot{a})$ then $u_a=0$,$u_{\dot{a}}=0$ and 
$u_{\ddot{a}}=\frac{\dot{u}}{\dddot{a}}$, and 
similarly for $v$. In this case  (\ref{FR1a}),(\ref{FR2a}) become a system of 
simple differential equations for 
$u(t)$ and $v(t)$, namely:
\begin{eqnarray}
&& h(u,v,\dot{u},\dot{v},t)=0\nonumber
\\
 &&g(u,v,\dot{u},\dot{v},\ddot{u},\ddot{v},t)=0.
\end{eqnarray} 
This system can be easily solved to find $u(t)$ and $v(t)$, and knowing also 
the imposed $a(t)$ we can eliminate time and reconstruct the pairs of 
$u(a)$,$v(a)$, or $u(\dot{a})$,$v(a)$, or $u(a)$,$v(\dot{a})$ etc. Hence, 
these are the connection forms that generate the imposed 
$H(t)$, which has the desired inflation phenomenology. We mention that the 
reconstruction of suitable $u$ and $v$, namely of a suitable connection, has 
equal right with the standard procedures of reconstructing the inflaton 
potential according to the desired inflationary observables. In the latter case 
one attributes the specific inflationary features to the unknown  physics 
(i.e. potential) of 
the inflaton field, while in the theory at hand the  inflationary features are 
attributed to a new, non-special, connection that determines the underlying 
geometry. The capability  of 
describing inflation solutions is an additional advantage of the theory at hand.

\subsection{$F(R,T)=R+\alpha R^2$}

In this subsection we are interested in examining a non-linear $F(R,T)$ 
case, in order for the  scalar degrees of freedom 
$\phi_1$,$\phi_2$ to be dynamical. We consider the simple case 
$F(R,T)=R+\alpha 
R^2$ and thus only $\phi_1$ switches on, while $\phi_2$ does not exist (the 
conformal transformation for $T$ is absent from first place). 
In this case we have that $F_{\phi_1}=1+2\alpha\phi_1$,  
 $V(\phi_1,\phi_2)=\alpha \phi_1^2$ and $f(\phi_1,\phi_2)=0$. Hence, equations
 (\ref{eq1})-(\ref{eq4}) reduce to
 \begin{eqnarray}
&&
\!\!\!\!\!\!\!\!\!\!\!\!\!\!\!\!\!\!\!
3H^{2}= 
\kappa^2\left[
\frac{\rho_m}{(1\!+\!2\alpha\phi_1)^2}  +\frac{1}{2}
\dot{\phi}^{2}_{1} + \alpha \phi_1^2
\right]
\nonumber\\
 &&
 +\left(\frac{Hau_{\dot{a}} -u}{2 }   
\right)+\frac{au_{\ddot{a}}}{2} 
\left(\dot{H}-2 H^2\right)
-\frac{1}{2} 
\dot{u}_{\ddot{a}}
,
\label{eq1c}
\end{eqnarray}
 \begin{eqnarray}
&&
\!\!\!\!\!\!\!\!\!\! \!\!\!\!\!\!\!\!\!\!\!\!\!\!\!\!\!
2\dot{H}+3H^2
=
\kappa^2
\left(-\frac{1}{2} 
\dot{\phi}^{2}_{1} + 
\alpha \phi_1^2\right)\nonumber\\
&& \ \ \ \
+
\left[\frac{Hau_{\dot{a}} -u}{2 }    -\frac{a}{6} 
\left(u_a-{\dot{u}_{\dot{a}}} \right)\right]\nonumber\\
&&\  \ \ \
+\frac{au_{\ddot{a}}}{2} 
 \left(\dot{H}+3H^2\right)+a H 
\dot{u}_{\ddot{a}}+\frac{a}{6} \ddot{u}_{\ddot{a}},
\label{eq2c}
\end{eqnarray}
 \begin{equation}
\ddot{\phi}_{1}=-3H\dot{\phi}_{1}-2\alpha\phi_1     
+\frac{4\alpha\rho_m}{(1+2\alpha\phi_1)^3}.
\label{eq3c}
\end{equation}
We can re-write the two Friedmann equations in the standard form 
(\ref{FR1a}),(\ref{FR2a}) defining
\begin{eqnarray}
&&
\!\!\!\!\!\!\!\!\!\!\!\!\!\!\!\!
\rho_{MG}=(1+2\alpha \phi_1)^2\Big[
\frac{1}{2} 
\dot{\phi}^{2}_{1} + 
\alpha \phi_1^2+  
\frac{Hau_{\dot{a}} -u }{2\kappa^2} \nonumber\\
&&
\ \ \ \ \ \ \ \ \ \ \ \ \ \ \ \ \ \ \,
+  
\frac{au_{\ddot{a}} 
\left(\dot{H}-2 H^2\right)
- 
\dot{u}_{\ddot{a}}}{2\kappa^2}   
\Big]\nonumber\\
&&\ \ \ \ \ \ \ \ \ \ \ \ \ \ \ \ \ \ \,
-\frac{12\alpha\phi_1}{\kappa^2}(1+\alpha\phi_1)H^2,
\label{rhoDEc1}\\
&&
\!\!\!\!\!\!\!\!\!\!\!\!\!\!\!\!
p_{MG}=\frac{1}{2} 
\dot{\phi}^{2}_{1} -\alpha \phi_1^2-\frac{1}{\kappa^2} \!
\Big[\frac{Hau_{\dot{a}}-u}{2}   -\frac{a}{6} 
\left(u_a\!-\!{\dot{u}_{\dot{a}}} \right)\nonumber\\
&&\ \ \ \ \ \ \ \  
+\frac{au_{\ddot{a}}}{2} 
 \left(\dot{H}+3H^2\right)+a H 
\dot{u}_{\ddot{a}}+\frac{a}{6} \ddot{u}_{\ddot{a}}
\Big].
\label{pDEc1}
\end{eqnarray} 

Let us now apply the above scenario at late times. In this case, the extra 
terms arising form the non-special connection will constitute an effective 
dark-energy sector with energy density and pressure given by  
(\ref{rhoDEc1}),(\ref{pDEc1}), and an equation-of-state parameter of the form 
\begin{eqnarray}
&&w_{MG}\equiv w_{DE}= \frac{p_{MG}}{\rho_{DE}}.
\label{wDEc}
\end{eqnarray}
\begin{figure}[ht]
\includegraphics[width=0.395\textwidth]{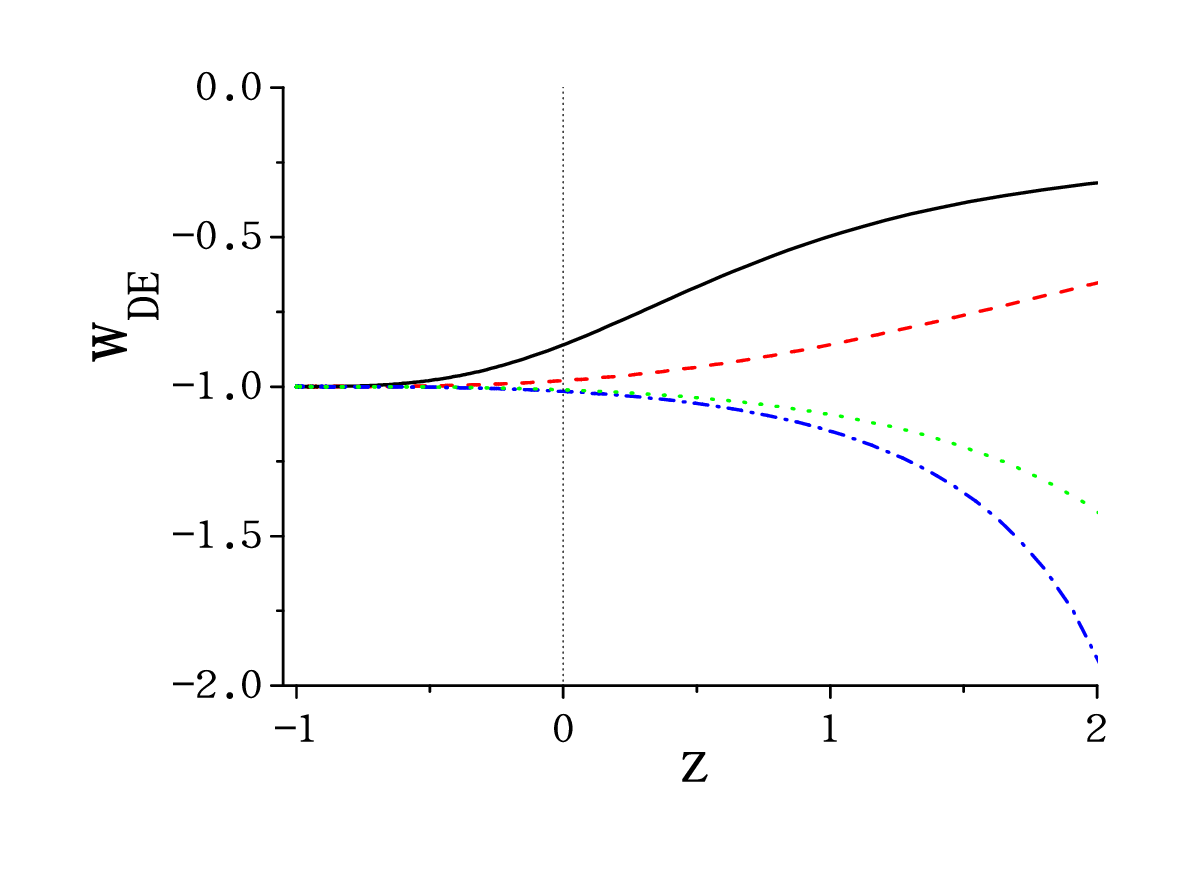}
\caption{
{\it{  The evolution of the equation-of-state parameter  $w_{DE}$  of the 
effective dark
energy    
in the case of  $F(R,T)=R+\alpha R^2$ and $u=c_1 \dot{a}-c_2$,  for various 
values of the model parameters:
 $\alpha=-0.2$,$c=1$ (black - solid),  $\alpha=-0.1$,$c=1$  (red - dashed), 
  $\alpha=0.05$,$c=1$  (green - dotted),  $\alpha=0.1$,$c=1$  (blue - 
dashed-dotted).
We have imposed the initial conditions
$\Omega_{DE}(z=0)\equiv\Omega_{DE0}\approx0.69$ and we use units where 
$\kappa^2=1$. The dotted vertical line marks the present time.}}}
\label{Model3awde}
\end{figure}
 Choosing $u=c_1 \dot{a}-c_2$ 
 with $c_1$,$c_2$ constants, we obtain
\begin{eqnarray}
&&
\!\!\!\!\!\!\!\!\!\!\!\!\!\!\!\!
\rho_{MG}=(1+2\alpha \phi_1)^2\left[
\frac{1}{2} 
\dot{\phi}^{2}_{1} + 
\alpha \phi_1^2+  
\frac{c_2}{\kappa^2}   
\right]\nonumber\\
&&\ \
-\frac{12\alpha\phi_1}{\kappa^2}(1+\alpha\phi_1)H^2,
\label{rhoDEc1b}\\
&&
\!\!\!\!\!\!\!\!\!\!\!\!\!\!\!\!
p_{MG}=\frac{1}{2} 
\dot{\phi}^{2}_{1} -\alpha \phi_1^2-\frac{c_2}{\kappa^2} .
\label{pDEc1b}
\end{eqnarray}
The above scenario can describe the sequence of epochs and one can obtain a 
cosmological evolution similar to the upper graph of Fig.  
\ref{Model1aomegas}. Additionally, in  Fig. \ref{Model3awde} we present the 
evolution of  $w_{DE}(z)$ for various parameter 
choices. As we observe $w_{DE}$ may be quintessence-like or phantom-like 
according to the parameter $\alpha$, and asymptotically it tends towards the 
cosmological constant value. We mention that $w_{DE}$ can lie in the phantom 
regime, despite the fact that the involved scalar field is canonical, namely the 
phantom behavior is induced due to the effect of the non-special connection. 
This feature shows the capabilities of the theory.

\section{Conclusions}
\label{Conclusions}

In the present work we investigated the cosmological applications of Myrzakulov 
$F(R,T)$ gravity. In this theory ones uses a specific but non-special 
connection, and thus both curvature and torsion are 
dynamical fields related to gravity. Thus, the theory has extra degrees 
of freedom arising from the non-special connection, as well as   extra degrees 
of freedom coming from the arbitrary function in the Lagrangian. 

In order to 
handle the non-special connection more efficiently, we introduced a 
parametrization that quantifies the deviation of curvature and torsion scalars 
form their corresponding values obtained using the special Levi-Civita 
and Weitzenb{\"{o}}ck  connections. Note that the connection is non-special but 
still it is a specific one, and thus only tetrad variation is needed. Finally, 
in order to bypass the complications of a general variation, we 
extracted the field equations following the mini-super-space procedure focusing 
on FRW geometry.

Considering the simple case where the action of the theory is linear in $R$ and 
$T$, namely $F(R,T)=R+\lambda T$, we found that, despite the fact that the 
action is simple, the Friedmann equations do contain new terms of geometrical 
origin,   reflecting the non-special connection.  Application in    
late-time universe gives rise to an effective dark-energy sector of geometrical 
origin. As we saw, we can obtain the 
thermal history of the universe, namely the sequence of matter and dark energy  
epochs, as required by observations. Moreover, we showed that the effective 
dark-energy 
equation-of-state parameter can be quintessence-like or phantom-like, or behave 
exactly as a cosmological constant and thus reproducing exactly $\Lambda$CDM 
cosmology. We mention that these features were 
obtained without the consideration of an  explicit cosmological constant,
namely 
they arise 
purely from the non-special connection and the intrinsic geometrical structure 
of the theory. These properties may be significant in the modified gravity 
model building.  Finally, early-time application   
showed that   we are able to extract a de Sitter 
solution, as well as to acquire more realistic inflation realizations with the 
desired  scale-factor evolution and thus with the desired 
values of the spectral-index and the tensor-to-scalar 
ratio, if we use the suitably reconstructed non-special connection.

Similarly, considering the case $F(R,T)=R+\alpha R^2$, in which one extra 
scalar degree of freedom appears on top of the effects of the non-special 
connection, we showed that we can also obtain the epoch sequence, and that the 
dark-energy equation-of-state parameter can be quintessence-like or phantom 
like, before tend asymptotically to the cosmological constant value. The 
interesting feature is that the phantom regime is obtained although the 
involved scalar field is canonical, which is an advantage, revealing the 
capabilities of the theory.

In summary, the rich structure of the underlying non-special connection leads 
the theory to have interesting cosmological phenomenology, both at early and 
late times. There are many additional studies that one should perform. 
One such interesting and necessary investigation is to  
   use data from  Type Ia Supernovae (SNIa), Cosmic 
Microwave Background (CMB) shift parameter, Baryon Acoustic Oscillations (BAO), 
and direct Hubble constant observations, in order to impose constraints on the 
model parameters and the involved non-special connection. An additional related 
study would be to examine the perturbations of the theory since this could both 
give information on the conditions for the absence of instabilities at the 
perturbative level, as well as to allow for a confrontation of the theory with 
perturbation related data such as   $f\sigma_8$ ones. The parametrization of 
the non-special connection through $u$ and $v$, that eventually depend on 
the tetrad and its derivatives, is a convenient framework for such an analysis 
since it bypasses complications. 
Furthermore, one can perform 
a detailed phase-space analysis in order to examine the global behavior of the 
theory independently from the initial 
conditions.
 Nevertheless,  such investigations, although both interesting and necessary, 
lie beyond the scope
of this work  and are left for future projects.

\begin{acknowledgments}
The work was carried out with the financial support of the Ministry of Education 
and Science of the Republic of Kazakhstan, Grant No. 0118RK00935. This article 
is based upon work from COST Action ``Cosmology and 
Astrophysics Network for
Theoretical Advances and Training Actions'', supported by COST (European 
Cooperation in
Science and Technology). 

\end{acknowledgments}


\begin{thebibliography}{99}

 
   %\cite{Capozziello:2011et}
\bibitem{Capozziello:2011et}
S.~Capozziello and M.~De Laurentis,
% {\it{Extended Theories of Gravity}},
Phys.\ Rept.\ {\bf 509}, 167 (2011)
[arXiv:1108.6266 [gr-qc]].



 
%\cite{Nojiri:2010wj}
\bibitem{Nojiri:2010wj} 
S.~Nojiri and S.~D.~Odintsov,
% {\it{Unified cosmic history in modified gravity: from F(R) theory to Lorentz 
%non-invariant models}},
Phys.\ Rept.\ {\bf 505}, 59 (2011)
[arXiv:1011.0544 [gr-qc]].


 



  %\cite{Copeland:2006wr}
\bibitem{Copeland:2006wr}
  E.~J.~Copeland, M.~Sami and S.~Tsujikawa,
  %``Dynamics of dark energy,''
  Int.\ J.\ Mod.\ Phys.\ D {\bf 15}, 1753 (2006)
  %doi:10.1142/S021827180600942X
  [hep-th/0603057].
  %%CITATION = doi:10.1142/S021827180600942X;%%
  %3758 citations counted in INSPIRE as of 19 Mar 2019
  
   
%\cite{Cai:2009zp}
\bibitem{Cai:2009zp} 
  Y.~F.~Cai, E.~N.~Saridakis, M.~R.~Setare and J.~Q.~Xia,
  %``Quintom Cosmology: Theoretical implications and observations,''
  Phys.\ Rept.\  {\bf 493}, 1 (2010)
 % doi:10.1016/j.physrep.2010.04.001
  [arXiv:0909.2776 [hep-th]].
  %%CITATION = doi:10.1016/j.physrep.2010.04.001;%%
  %535 citations counted in INSPIRE as of 23 Jul 2019

%\cite{Bartolo:2004if}
\bibitem{Bartolo:2004if} 
N.~Bartolo, E.~Komatsu, S.~Matarrese and A.~Riotto,
% {\it{Non-Gaussianity from inflation: Theory and observations}},
Phys.\ Rept.\ {\bf 402}, 103 (2004)
[astro-ph/0406398].


 
 %\cite{Stelle:1976gc}
\bibitem{Stelle:1976gc} 
  K.~S.~Stelle,
  %``Renormalization of Higher Derivative Quantum Gravity,''
  Phys.\ Rev.\ D {\bf 16}, 953 (1977).
  %doi:10.1103/PhysRevD.16.953
  %%CITATION = doi:10.1103/PhysRevD.16.953;%%
  %1758 citations counted in INSPIRE as of 05 Dec 2019
  
%\cite{Biswas:2011ar}
\bibitem{Biswas:2011ar} 
  T.~Biswas, E.~Gerwick, T.~Koivisto and A.~Mazumdar,
  %``Towards singularity and ghost free theories of gravity,''
  Phys.\ Rev.\ Lett.\  {\bf 108}, 031101 (2012)
  %doi:10.1103/PhysRevLett.108.031101
  [arXiv:1110.5249 [gr-qc]].
  %%CITATION = doi:10.1103/PhysRevLett.108.031101;%%
  %386 citations counted in INSPIRE as of 05 Dec 2019

 

  %\cite{DeFelice:2010aj}
\bibitem{DeFelice:2010aj}
  A.~De Felice and S.~Tsujikawa,
  %``f(R) theories,''
  Living Rev.\ Rel.\  {\bf 13}, 3 (2010).
%   [arXiv:1002.4928 [gr-qc]].
  %%CITATION = ARXIV:1002.4928;%%

 



%\cite{Nojiri:2005jg}
\bibitem{Nojiri:2005jg}
  S.~'i.~Nojiri and S.~D.~Odintsov,
  %``Modified Gauss-Bonnet theory as gravitational alternative for dark
%energy,''
  Phys.\ Lett.\ B {\bf 631}, 1 (2005).
%  [hep-th/0508049].
  %%CITATION = HEP-TH/0508049;%%
  %222 citations counted in INSPIRE as of 14 Feb 2014


  %\cite{DeFelice:2008wz}
\bibitem{DeFelice:2008wz}
  A.~De Felice and S.~Tsujikawa,
  %``Construction of cosmologically viable f(G) dark energy models,''
  Phys.\ Lett.\ B {\bf 675}, 1 (2009).
%  [arXiv:0810.5712 [hep-th]].
  %%CITATION = ARXIV:0810.5712;%%
  %69 citations counted in INSPIRE as of 14 Feb 2014


 %\cite{Lovelock:1971yv}
\bibitem{Lovelock:1971yv}
  D.~Lovelock,
  %``The Einstein tensor and its generalizations,''
  J.\ Math.\ Phys.\  {\bf 12}, 498 (1971).
  %%CITATION = JMAPA,12,498;%%
  %824 citations counted in INSPIRE as of 14 Feb 2014


  %\cite{Deruelle:1989fj}
\bibitem{Deruelle:1989fj}
  N.~Deruelle and L.~Farina-Busto,
  %``The Lovelock Gravitational Field Equations in Cosmology,''
  Phys.\ Rev.\ D {\bf 41}, 3696 (1990).
  %%CITATION = PHRVA,D41,3696;%%
  %78 citations counted in INSPIRE as of 14 Feb 2014

\bibitem{Pereira}
R. Aldrovandi and J. G. Pereira, {\it Teleparallel Gravity: An Introduction},
Springer, Dordrecht (2013).


  %\cite{Maluf:2013gaa}
\bibitem{Maluf:2013gaa}
  J.~W.~Maluf,
  %``The teleparallel equivalent of general relativity,''
  Annalen Phys.\  {\bf 525}, 339 (2013).
  %%CITATION = ARXIV:1303.3897;%%
  %10 citations counted in INSPIRE as of 14 Feb 2014



  
  
  
  
  %\cite{Cai:2015emx}
\bibitem{Cai:2015emx}
  Y.~F.~Cai, S.~Capozziello, M.~De Laurentis and E.~N.~Saridakis,
  %``f(T) teleparallel gravity and cosmology,''
  Rept.\ Prog.\ Phys.\  {\bf 79}, no. 10, 106901 (2016)
% % doi:10.1088/0034-4885/79/10/106901
 [arXiv:1511.07586 [gr-qc]].
  %%CITATION =% doi:10.1088/0034-4885/79/10/106901;%%
  %137 citations counted in INSPIRE as of 06 Oct 2017
  
  
  
  
%\cite{Ferraro:2006jd}
\bibitem{Ferraro:2006jd}
  R.~Ferraro and F.~Fiorini,
  %``Modified teleparallel gravity: Inflation without inflaton,''
  Phys.\ Rev.\ D {\bf 75}, 084031 (2007)
  [gr-qc/0610067].
 

%\cite{Linder:2010py}
\bibitem{Linder:2010py} 
  E.~V.~Linder,
  %``Einstein's Other Gravity and the Acceleration of the Universe,''
  Phys.\ Rev.\ D {\bf 81}, 127301 (2010)
 % Erratum: [Phys.\ Rev.\ D {\bf 82}, 109902 (2010)]
  %doi:10.1103/PhysRevD.81.127301, 10.1103/PhysRevD.82.109902
  [arXiv:1005.3039 [astro-ph.CO]].
  %%CITATION = doi:10.1103/PhysRevD.81.127301, 10.1103/PhysRevD.82.109902;%%
  %558 citations counted in INSPIRE as of 05 Dec 2019
  
    %\cite{Kofinas:2014owa}
\bibitem{Kofinas:2014owa} 
  G.~Kofinas and E.~N.~Saridakis,
  %``Teleparallel equivalent of Gauss-Bonnet gravity and its modifications,''
  Phys.\ Rev.\ D {\bf 90}, 084044 (2014)
  %doi:10.1103/PhysRevD.90.084044
  [arXiv:1404.2249 [gr-qc]].
  %%CITATION =% doi:10.1103/PhysRevD.90.084044;%%
  %116 citations counted in INSPIRE as of 11 Jul 2019
  
  
  
  %\cite{Harko:2018gxr}
\bibitem{Harko:2018gxr} 
  T.~Harko, T.~S.~Koivisto, F.~S.~N.~Lobo, G.~J.~Olmo and D.~Rubiera-Garcia,
  %``Coupling matter in modified $Q$ gravity,''
  Phys.\ Rev.\ D {\bf 98}, no. 8, 084043 (2018)
  %doi:10.1103/PhysRevD.98.084043
  [arXiv:1806.10437 [gr-qc]].
  %%CITATION = doi:10.1103/PhysRevD.98.084043;%%
  %17 citations counted in INSPIRE as of 05 Dec 2019

    

  
  
  
%\cite{Bogoslovsky:1999pp}
\bibitem{Bogoslovsky:1999pp} 
  G.~Y.~Bogoslovsky and H.~F.~Goenner,
  % {\it{Finslerian spaces possessing local relativistic symmetry}},
  Gen.\ Rel.\ Grav.\  {\bf 31}, 1565 (1999)
%  doi:10.1023/A:1026786505326
  [gr-qc/9904081].
  %%CITATION = doi:10.1023/A:1026786505326;%%
  %33 citations counted in INSPIRE as of 10 Feb 2019
  
%\cite{Mavromatos:2010jt}
\bibitem{Mavromatos:2010jt} 
  N.~E.~Mavromatos, S.~Sarkar and A.~Vergou,
 %{\it{Stringy Space-Time Foam, Finsler-like Metrics and Dark Matter Relics}},
  Phys.\ Lett.\ B {\bf 696}, 300 (2011)
%  doi:10.1016/j.physletb.2010.12.045
  [arXiv:1009.2880 [hep-th]].
  %%CITATION = doi:10.1016/j.physletb.2010.12.045;%%
  %27 citations counted in INSPIRE as of 10 Feb 2019
  %\cite{Basilakos:2013hua}

 \bibitem{kour-stath-st 2012}
  A.~P.~Kouretsis, M.~Stathakopoulos and P.~C.~Stavrinos,
  %``Covariant kinematics and gravitational bounce in Finsler space-times,''
  Phys.\ Rev.\ D {\bf 86}, 124025 (2012)
  %doi:10.1103/PhysRevD.86.124025
  [arXiv:1208.1673 [gr-qc]].
  %%CITATION = doi:10.1103/PhysRevD.86.124025;%%
  %23 citations counted in INSPIRE as of 23 Jul 2019
  
  
  \bibitem{Basilakos:2013hua} 
  S.~Basilakos, A.~P.~Kouretsis, E.~N.~Saridakis and P.~Stavrinos,
  %{\it{Resembling dark energy and modified gravity with Finsler-Randers 
%cosmology}},
  Phys.\ Rev.\ D {\bf 88}, 123510 (2013)
%  doi:10.1103/PhysRevD.88.123510
  [arXiv:1311.5915 [gr-qc]].
  %%CITATION = doi:10.1103/PhysRevD.88.123510;%%
  %29 citations counted in INSPIRE as of 10 Feb 2019
  
  
  
  
  
  %\cite{Triantafyllopoulos:2018bli}
\bibitem{Triantafyllopoulos:2018bli} 
  A.~Triantafyllopoulos and P.~C.~Stavrinos,
 %\emph{Weak field equations and generalized FRW cosmology on the tangent 
%Lorentz bundle},
  Class.\ Quant.\ Grav.\  {\bf 35}, no. 8, 085011 (2018).
%  doi:10.1088/1361-6382/aab27f
  %%CITATION = doi:10.1088/1361-6382/aab27f;%%
  
  
  %\cite{Ikeda:2019ckp}
\bibitem{Ikeda:2019ckp} 
  S.~Ikeda, E.~N.~Saridakis, P.~C.~Stavrinos and A.~Triantafyllopoulos,
  %``Cosmology of Lorentz fiber-bundle induced scalar-tensor theories,''
  arXiv:1907.10950 [gr-qc].
  %%CITATION = ARXIV:1907.10950;%%
  %1 citations counted in INSPIRE as of 05 Dec 2019

  %\cite{Minas:2019urp}
\bibitem{Minas:2019urp} 
  G.~Minas, E.~N.~Saridakis, P.~C.~Stavrinos and A.~Triantafyllopoulos,
  %``Bounce cosmology in generalized modified gravities,''
  Universe {\bf 5}, 74 (2019)
  %doi:10.3390/universe5030074
  [arXiv:1902.06558 [gr-qc]].
  %%CITATION = doi:10.3390/universe5030074;%%
  %2 citations counted in INSPIRE as of 06 Dec 2019

%\cite{Hehl:1994ue}
\bibitem{Hehl:1994ue} 
  F.~W.~Hehl, J.~D.~McCrea, E.~W.~Mielke and Y.~Ne'eman,
  %``Metric affine gauge theory of gravity: Field equations, Noether 
identities, 
%world spinors, and breaking of dilation invariance,''
  Phys.\ Rept.\  {\bf 258}, 1 (1995)
 % doi:10.1016/0370-1573(94)00111-F
  [gr-qc/9402012].
  %%CITATION = doi:10.1016/0370-1573(94)00111-F;%%
  %981 citations counted in INSPIRE as of 05 Dec 2019
  
%\cite{BeltranJimenez:2012sz}
\bibitem{BeltranJimenez:2012sz} 
  J.~Beltran Jimenez, A.~Golovnev, M.~Karciauskas and T.~S.~Koivisto,
  %``The Bimetric variational principle for General Relativity,''
  Phys.\ Rev.\ D {\bf 86}, 084024 (2012)
 % doi:10.1103/PhysRevD.86.084024
  [arXiv:1201.4018 [gr-qc]].
  %%CITATION = doi:10.1103/PhysRevD.86.084024;%%
  %25 citations counted in INSPIRE as of 05 Dec 2019

  
  
   %\cite{Tamanini:2012mi}
\bibitem{Tamanini:2012mi} 
  N.~Tamanini,
  %``Variational approach to gravitational theories with two independent 
%connections,''
  Phys.\ Rev.\ D {\bf 86}, 024004 (2012)
 % doi:10.1103/PhysRevD.86.024004
  [arXiv:1205.2511 [gr-qc]].
  %%CITATION = doi:10.1103/PhysRevD.86.024004;%%
  %13 citations counted in INSPIRE as of 05 Dec 2019
  
  
  
%\cite{Myrzakulov:2012qp}
\bibitem{Myrzakulov:2012qp} 
  R.~Myrzakulov,
  %``FRW Cosmology in F(R,T) gravity,''
  Eur.\ Phys.\ J.\ C {\bf 72}, 2203 (2012)
 %doi:10.1140/epjc/s10052-012-2203-y
  [arXiv:1207.1039 [gr-qc]].
  %%CITATION =%doi:10.1140/epjc/s10052-012-2203-y;%%
  %88 citations counted in INSPIRE as of 03 Dec 2019


  
%\cite{Harko:2011kv}
\bibitem{Harko:2011kv} 
  T.~Harko, F.~S.~N.~Lobo, S.~Nojiri and S.~D.~Odintsov,
  %``$f(R,T)$ gravity,''
  Phys.\ Rev.\ D {\bf 84}, 024020 (2011)
 % doi:10.1103/PhysRevD.84.024020
  [arXiv:1104.2669 [gr-qc]].
  %%CITATION = doi:10.1103/PhysRevD.84.024020;%%
  %730 citations counted in INSPIRE as of 03 Dec 2019
  
  
  %\cite{Conroy:2017yln}
\bibitem{Conroy:2017yln} 
  A.~Conroy and T.~Koivisto,
  %``The spectrum of symmetric teleparallel gravity,''
  Eur.\ Phys.\ J.\ C {\bf 78}, no. 11, 923 (2018)
  %doi:10.1140/epjc/s10052-018-6410-z
  [arXiv:1710.05708 [gr-qc]].
  %%CITATION = doi:10.1140/epjc/s10052-018-6410-z;%%
  %30 citations counted in INSPIRE as of 18 Dec 2019

  
  
%\cite{Sharif:2012gz}
\bibitem{Sharif:2012gz} 
  M.~Sharif, S.~Rani and R.~Myrzakulov,
  %``Analysis of $F(R,T)$ gravity models through energy conditions,''
  Eur.\ Phys.\ J.\ Plus {\bf 128}, 123 (2013)
 %doi:10.1140/epjp/i2013-13123-0
  [arXiv:1210.2714 [gr-qc]].
  %%CITATION =%doi:10.1140/epjp/i2013-13123-0;%%
  %39 citations counted in INSPIRE as of 03 Dec 2019

 

%\cite{Momeni:2011am}
\bibitem{Momeni:2011am} 
  M.~Jamil, D.~Momeni, M.~Raza and R.~Myrzakulov,
  %``Reconstruction of some cosmological models in f(R,T) gravity,''
  Eur.\ Phys.\ J.\ C {\bf 72}, 1999 (2012).
 % doi:10.1140/epjc/s10052-012-1999-9
%  [arXiv:1107.5807 [physics.gen-ph]].
  %%CITATION = doi:10.1140/epjc/s10052-012-1999-9;%%
  %195 citations counted in INSPIRE as of 04 Dec 2019


%\cite{Capozziello:2014bna}
\bibitem{Capozziello:2014bna} 
  S.~Capozziello, M.~De Laurentis and R.~Myrzakulov,
  %``Noether Symmetry Approach for teleparallel-curvature cosmology,''
  Int.\ J.\ Geom.\ Meth.\ Mod.\ Phys.\  {\bf 12}, no. 09, 1550095 (2015)
 %doi:10.1142/S0219887815500954
  [arXiv:1412.1471 [gr-qc]].
  %%CITATION =%doi:10.1142/S0219887815500954;%%
  %19 citations counted in INSPIRE as of 03 Dec 2019


 
%\cite{Feola:2019zqg}
\bibitem{Feola:2019zqg} 
  P.~Feola, X.~J.~Forteza, S.~Capozziello, R.~Cianci and S.~Vignolo,
  %``The mass-radius relation for neutron stars in $f(R)=R+\lambda R^2$ 
%gravity: 
%a comparison between purely metric and torsion formulations,''
  arXiv:1909.08847 [astro-ph.HE].
  %%CITATION = ARXIV:1909.08847;%%

  %\cite{Paliathanasis:2014iva}
\bibitem{Paliathanasis:2014iva} 
  A.~Paliathanasis, S.~Basilakos, E.~N.~Saridakis, S.~Capozziello, K.~Atazadeh, 
F.~Darabi and M.~Tsamparlis,
  %``New Schwarzschild-like solutions in f(T) gravity through Noether 
%symmetries,''
  Phys.\ Rev.\ D {\bf 89}, 104042 (2014)
 % doi:10.1103/PhysRevD.89.104042
  [arXiv:1402.5935 [gr-qc]].
  %%CITATION = doi:10.1103/PhysRevD.89.104042;%%
  %97 citations counted in INSPIRE as of 09 Dec 2019

  
   %\cite{Paliathanasis:2015aos}
\bibitem{Paliathanasis:2015aos} 
  A.~Paliathanasis,
  %``$f(R)$-gravity from Killing Tensors,''
  Class.\ Quant.\ Grav.\  {\bf 33}, no. 7, 075012 (2016)
 % doi:10.1088/0264-9381/33/7/075012
  [arXiv:1512.03239 [gr-qc]].
  %%CITATION = doi:10.1088/0264-9381/33/7/075012;%%
  %38 citations counted in INSPIRE as of 05 Dec 2019

  
 %\cite{Dimakis:2016mip}
\bibitem{Dimakis:2016mip} 
  N.~Dimakis, A.~Karagiorgos, A.~Zampeli, A.~Paliathanasis, T.~Christodoulakis 
and P.~A.~Terzis,
  %``General Analytic Solutions of Scalar Field Cosmology with Arbitrary 
%Potential,''
  Phys.\ Rev.\ D {\bf 93}, no. 12, 123518 (2016)
 % doi:10.1103/PhysRevD.93.123518
  [arXiv:1604.05168 [gr-qc]].
  %%CITATION = doi:10.1103/PhysRevD.93.123518;%%
  %20 citations counted in INSPIRE as of 05 Dec 2019

  
  

%\cite{Yang:2010ji}
\bibitem{Yang:2010ji} 
  R.~J.~Yang,
  %``Conformal transformation in $f(T)$ theories,''
  EPL {\bf 93}, no. 6, 60001 (2011)
 %doi:10.1209/0295-5075/93/60001
  [arXiv:1010.1376 [gr-qc]].
  %%CITATION =%doi:10.1209/0295-5075/93/60001;%%
  %152 citations counted in INSPIRE as of 03 Dec 2019
  
  %\cite{Wright:2016ayu}
\bibitem{Wright:2016ayu} 
  M.~Wright,
  %``Conformal transformations in modified teleparallel theories of gravity 
%revisited,''
  Phys.\ Rev.\ D {\bf 93}, no. 10, 103002 (2016)
 %doi:10.1103/PhysRevD.93.103002
  [arXiv:1602.05764 [gr-qc]].
  %%CITATION =%doi:10.1103/PhysRevD.93.103002;%%
  %28 citations counted in INSPIRE as of 03 Dec 2019
  

  
%\cite{Gakis:2019rdd}
\bibitem{Gakis:2019rdd} 
  V.~Gakis, M.~Kr\v{s}\v{s}\'ak, J.~Levi Said and E.~N.~Saridakis,
  %``Conformal Gravity and Transformations in the Symmetric Teleparallel 
%Framework,''
  arXiv:1908.05741 [gr-qc].
  %%CITATION = ARXIV:1908.05741;%%
 
 
  
  
  %\cite{Ade:2015xua}
\bibitem{Ade:2015xua}
  P.~A.~R.~Ade {\it et al.} [Planck Collaboration],
 %``Planck 2015 results. XIII. Cosmological parameters,''
  Astron.\ Astrophys.\  {\bf 594}, A13 (2016)
%  doi:10.1051/0004-6361/201525830
  [arXiv:1502.01589 [astro-ph.CO]].
  %%CITATION = doi:10.1051/0004-6361/201525830;%%
  %6704 citations counted in INSPIRE as of 07 Feb 2019
  
  
  
\bibitem{Martin:2013tda}
  J.~Martin, C.~Ringeval and V.~Vennin,
  %``Encyclopædia Inflationaris,''
  Phys.\ Dark Univ.\  {\bf 5-6} (2014) 75
 % doi:10.1016/j.dark.2014.01.003
  [arXiv:1303.3787 [astro-ph.CO]].

  
\end{thebibliography}
\end{document}